\documentclass{mn2e}
\usepackage{epsfig,natbib2,natbibmnfix}
\usepackage{graphicx}
\usepackage{amssymb}
\usepackage{multirow}
\usepackage{multicol}
\usepackage{longtable}
\usepackage[varg]{txfonts}
\usepackage{color}
\usepackage{url}
\newcommand{\cref}[1]{Chapter~\ref{#1}}

\begin{document}
%%%%%%%%%%%%%%%%%%%%%%%%%%%%%%%%%%%%%%%%%%%%%%%%%%%%%%%%%%%%%%%%%%%%%%%%%%%%%%
%% Title Details and Page Header                                            %%
%%%%%%%%%%%%%%%%%%%%%%%%%%%%%%%%%%%%%%%%%%%%%%%%%%%%%%%%%%%%%%%%%%%%%%%%%%%%%%
\title[A long-term variable hot DQ]{SDSS J000555.90$-$100213.5: a hot, magnetic carbon-atmosphere WD rotating with a 2.1 day period} \author[K.~A.~Lawrie et al.] {
\parbox{7in}{K. A. Lawrie$^1$\footnotemark[1], M. R. Burleigh$^1$, P. Dufour$^2$ \& S. T. Hodgkin$^3$}
\vspace{0.1in}
  \\ $^1$ Department of Physics \& Astronomy, University of Leicester, Leicester, LE1 7RH, UK
  \\ $^2$ D{\'e}partement de Physique, Universit{\'e} de Montr{\'e}al, C.P. 6128, Succ. Centre-Ville, Montr{\'e}al, QC H3C 3J7, Canada
  \\ $^3$ Cambridge Astronomical Survey Unit, Institute of Astronomy, Madingley Road, Cambridge, CB3 0HA, UK
 }

\maketitle

%%%%%%%%%%%%%%%%%%%%%%%%%%%%%%%%%%%%%%%%%%%%%%%%%%%%%%%%%%%%%%%%%%%%%%%%%%%%%%
%% Abstract, Keywords and contact details                                   %%
%%%%%%%%%%%%%%%%%%%%%%%%%%%%%%%%%%%%%%%%%%%%%%%%%%%%%%%%%%%%%%%%%%%%%%%%%%%%%%
\begin{abstract}
A surprisingly large fraction (70\%) of hot, carbon dominated atmosphere (DQ) white dwarfs are magnetic and/or photometrically variable on short timescales up to $\sim$1000 s. However, here we show that the hot DQ magnetic white dwarf SDSS J000555.90-100213.5 is photometrically variable by 11\% on a longer timescale, with a period of $2.110\pm0.045$ days. We find no evidence of the target fluctuating on short timescales at an amplitude of $\lesssim$$\pm0.5\%$. Short period hot DQ white dwarfs have been interpreted as non-radial pulsators, but in the case of SDSS J0005-1002, it is more likely that the variability is due to the magnetic hot DQ white dwarf rotating. We suggest that some hot DQ white dwarfs, varying on short timescales, should be more carefully examined to ascertain whether the variability is due to rotation rather than pulsation. All hot DQs should be monitored for long period modulations as an indicator of rotation and magnetism. 
\end{abstract}


\footnotetext[1]{E-mail: kal27@le.ac.uk}

%%%%%%%%%%%%%%%%%%%%%%%%%%%%%%%%%%%%%%%%%%%%%%%%%%%%%%%%%%%%%%%%%%%%%%%%%%%%%%
%% Introduction                                                             %%
%%%%%%%%%%%%%%%%%%%%%%%%%%%%%%%%%%%%%%%%%%%%%%%%%%%%%%%%%%%%%%%%%%%%%%%%%%%%%%
\section{Introduction}
\label{intro}

Hot DQ white dwarfs have atmospheres dominated by carbon, containing little or no hydrogen or helium (\citealt{Dufour07}; \citealt{Dufour08b}). Only 14 hot DQs have been discovered so far, making them a rare class of white dwarf (\citealt{Dufour10}; \citealt{Liebert03}). In addition, their effective temperatures appear to cover a very specific range of $18,000-24,000$ K (\citealt{Dufour08b}). Zeeman split lines, indicative of the presence of magnetic fields, have been detected, or at least suspected, in 10 of the 14 catalogued hot DQs (70\%, \citealt{Dufour10}; \citealt{Dufour13}). In contrast, the fraction of magnetic white dwarfs in the general white dwarf population is thought to be in the range of $3-15$\% (\citealt{Kepler13}; \citealt{Jordan07}; \citealt{Landstreet12}; \citealt{Lieb03}), suggesting that perhaps all hot DQs are magnetic. 

\cite{Montgomery08} observed six hot DQ white dwarfs for pulsations and discovered the first photometrically variable hot DQ, SDSS J142625.71-575218.3 (hereafter SDSS J1426-5752), with modes at 417.7 s and 208.8 s (the first harmonic). Their theoretical calculations predicted that SDSS J1426-5752 should be the only star in the sample to pulsate, as it was the nearest to the high-temperature boundary (the ``blue edge") of the DQ white dwarf instability strip. Since then, however, variability has been detected for a further four hot DQ white dwarfs (SDSS J2200-0741 and SDSS J2348-0942, \citealt{Barlow08}; SDSS J1337-0026, \citealt{Dunlap10} and SDSS J1153+0056, \citealt{Dufour11}), the latter detected in the FUV using the Hubble Space Telescope (HST) and Cosmic Origins Spectrograph (COS), where the amplitudes of the modes were 2--4 times larger than those observed in the optical.

Here, we introduce the sixth variable hot DQ, SDSS J$000555.90-100213.5$ (hereafter SDSS J0005-1002). It was first discovered as a possible magnetic DQ white dwarf in SDSS DR1 (\citealt{Schmidtetal03}). It has the largest mean field strength for a hot DQ at 1.47 MG, measured from the line splitting in its spectrum (\citealt{Dufour08b}), and has an effective surface temperature of 19,420\,K. As part of a survey of magnetic white dwarfs searching for photometric variability (Lawrie et al. 2013, in prep), we detect modulations for SDSS J0005-1002 on the timescale of days. This periodicity is much longer than has been observed for the other hot DQ variables, which have thus far been interpreted as pulsations (\citealt{Barlow08}; \citealt{Dunlap10}; \citealt{Dufour11}). We discuss the cause of variability in SDSS J0005-1002, the photometric variability of hot DQ stars in general, the unusual pulsations some of them display, and the role of magnetism and its possible influence. 

%%%%%%%% SECTION 2 %%%%%%%%%%%%%%%%%%%%%%%%%%%%%%%%%%%%%%%%%%%%%%%%%
\section{Observations \& Data Reduction}

\begin{table}
\caption{Observations log of SDSS J0005-1002.}\label{tab:obs_log}
\begin{tabular}{llcccl}
\hline
\hline
Telescope & UT Date & Start Time & $T_{exp}$ & N   & Filter\\
                    &                 & (UTC)        & (s)                &      & \\      
\hline
INT WFC                 & 2009-10-17 &  23:03:42  & 60    &   3    & r$^{\prime}$ \\ % to 23:08:18 length=276s
INT WFC                 & 2009-10-18 &  21:15:48  & 120  &   9    & r$^{\prime}$ \\ %to 00:51:38 length=1368s
INT WFC                 & 2009-10-21 &  23:28:50  & 120  &  12  & r$^{\prime}$ \\% to 00:01:42 length=1972s
INT WFC                 & 2009-10-23 &  01:23:23  & 120  &   3   & r$^{\prime}$ \\% to 01:31:00 length=457s
SAAO 1.0m STE3 & 2012-09-08 &  21:09:28 & 180   &  75  & none \\ %to 00:53:33 length=13445s
SAAO 1.0m STE3 & 2012-09-10 &  22:08:12 & 180   &  54  & none \\ %to 00:50:08 length=9716s
SAAO 1.0m STE3 & 2012-09-11 &  22:04:36 & 180   &  55  & none \\ %to 00:52:58 length=10102s
SAAO 1.0m STE3 & 2012-10-18 &  20:24:41 & 180   &  13  & none \\ %to 21:05:48 N=13
SAAO 1.0m STE3 & 2012-10-22 &  18:57:44 & 90     &   29 & none \\ %to 20:18:33 N=29
SAAO 1.0m STE3 & 2012-10-23 &  18:46:04 & 120   &   60 & none \\ %to 20:54:46 N=60
\hline
\end{tabular}
\end{table}

\subsection{INT Optical Photometry}\label{sec:int_data}

SDSS J0005-1002 was observed using the 2.5\,m Isaac Newton Telescope (INT) in La Palma during a run from $17-23$ October 2009. The Wide Field Camera (WFC) was mounted on the INT at prime focus, and is made up of a mosaic of four 2k$\times$4k pixel CCDs with a total field-of-view of 34$\times$34\,arcmin$^2$ and a pixel scale of 0.33\,arcsec/pixel. Details are listed in an observations log in Table \ref{tab:obs_log}.

The data reduction was carried out using a pipeline developed by the Cambridge Astronomical Survey Unit (CASU); a detailed description of the process can be found in \cite{Irwin01} and \cite{Irwin07}. The pipeline performed a standard CCD reduction of bias correction, trimming of the frames, non-linearity correction, flat-fielding and gain correction. This was followed by an astrometric calibration of each frame, where the point source catalogue (PSC) from the Two-Micron All-Sky Survey (2MASS) was used as a reference astrometric catalogue. To get the optimal positions for the stars and thus reduce the positioning error, the aperture positions were determined by accurately finding the relative centroid positions of all of stars in the frame. This was carried out by stacking ten frames for each target field (taken in the best seeing and sky conditions) to create a master frame, giving a master catalogue listing all of the sources and coordinates in the image. The master frame was then used to determine the respective positions in the individual frames in the time series. 

The background level was determined by dividing the image into a coarse grid, where the clipped median of the counts for each bin was estimated (bad pixels were rejected using confidence maps). For a given pixel in the image, the background level was then calculated using bilinear interpolation over the background grid. This technique has been discussed in more detail in \cite{Irwin85}.

For the aperture photometry, the flux (and thus light curve) for each star was initially calculated for a range of increasing aperture radii ($r_{\rm core}$/2, $r_{\rm core}$, $\sqrt{2}\,r_{\rm core}$, 2$r_{\rm core}$ and $2\sqrt{2}\,r_{\rm core}$, where $r_{\rm core}$ was set to the typical FWHM and kept fixed for all of the data). The aperture size that yielded the smallest root-mean-square (rms) was chosen. Aperture corrections were then used to account for the different amounts of flux due to the differently-sized apertures, allowing for the same zero-point system to be used for all of the stars. To produce light curves for each of the stars, differential photometry was performed by calculating the flux for each star with respect to all of the stable stars in the field-of-view. The flux measurements were then converted to magnitudes using the zero-point estimate. Fluctuations in the photometry due to atmospheric effects, such as variations in transparency and extinction, were removed by fitting a 2D polynomial to the magnitude residuals of each non-variable star in the field to determine a zero-point correction. The photometric errors were calculated as the quadrature sum of the Poisson noise in the object's counts, Poisson noise in the sky, rms of the sky background fit and a constant value of $\sim$1.5\,mmag to account for systematic errors. The photometry and light curve production was discussed in \cite{Irwin07}.

\subsection{SAAO 1.0 m Optical Photometry}

We also observed SDSS J0005-1002 with the 1.0 m telescope located at the South African Astronomical Observatory (SAAO) from 29 August -- 11 September 2012 and $17-23$ October 2012. We used the SAAO CCD (STE3) instrument on the 1.0 m, which has a field-of-view of $512\times512$ pixels and a pixel scale of 0.31\,arcsec/pixel. Observation details are given in Table \ref{tab:obs_log}. 

\begin{figure}
\begin{center}
\includegraphics[width=8cm, trim=0cm 4cm 0cm 4cm, clip]{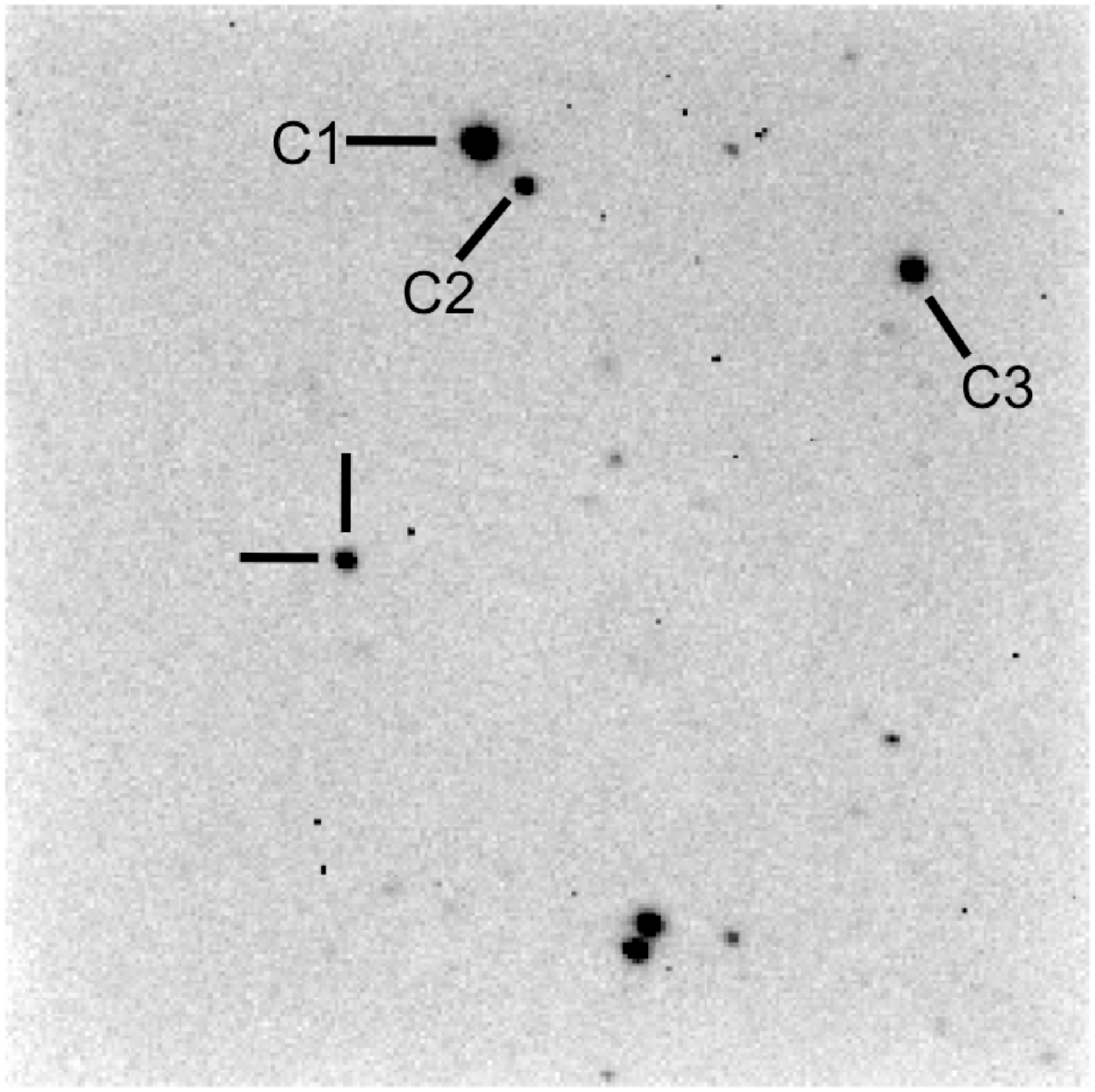}
\caption{Finder chart for SDSS J0005-1002, showing the SAAO STE3 CCD field-of-view (2.6'$\times$2.6'), where the target is marked by two bars and the comparison stars by C1, C2 and C3. North is towards the top of the frame and east is to the left.}\label{fig:sdss0005_saao_finderchart}
\end{center}
\end{figure}

The data were reduced using the SAAO CCD pipeline, which subtracted the bias and normalised by the master flat field frame. We used the \textsc{starlink} package \textsc{autophotom} to perform the photometry of the target and comparison stars. Figure \ref{fig:sdss0005_saao_finderchart} shows the SAAO STE3 field-of-view with the stars marked accordingly. The aperture width was fixed for a given night and was defined as 1.5 times the mean seeing (FWHM, \citealt{Naylor98}). This aperture size limited the contamination of background noise in the aperture, which was high due to a significant amount of moonlight during the September run. The sky background level was determined using the clipped mean of the pixel values in an annulus around the stars and the measurement errors were estimated from the sky variance. To remove atmospheric fluctuations, the light curve of SDSS J0005-1002 was divided by the light curve of one of the comparison stars. 

%\begin{table}

%\end{table}

%%%%%%%%%%% SECTION 3 %%%%%%%%%%%%%%%%%%%%%%%%%%%%%%%%%%%%%%%%%%%%%%%

\section{Analysis \& Results}

All time stamps are converted to the barycentric Julian date (BJD) using an IDL implementation by \cite*{Eastman10}. To assess the periodicity of the light curves, we use two different methods: a Fourier analysis using Period04 (\citealt{Lenz05}) and a least-squares fit of a sinusoid using MPFIT in IDL (\citealt{Markwardt09}). 

\begin{figure}
\begin{center}
\includegraphics[width=8cm]{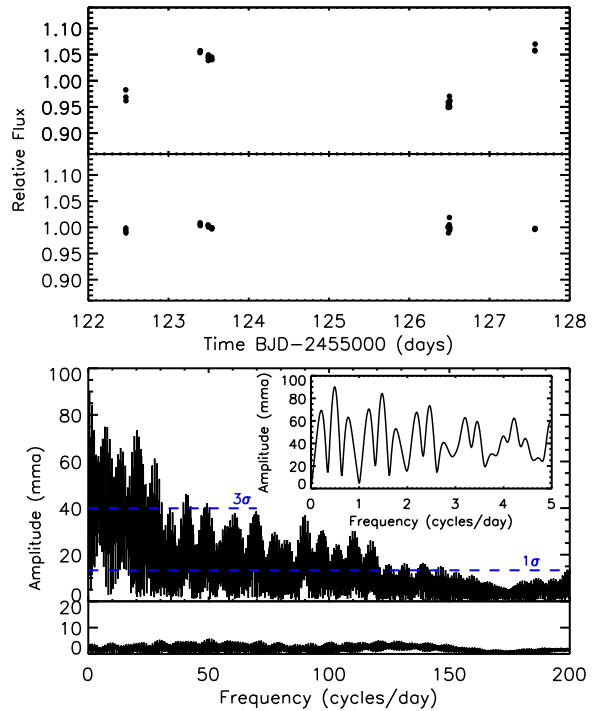}
\caption{\textit{Top of upper panel:} Differential light curve of SDSS J0005-1002 (target/(comp1+comp3)) taken using the INT WFC in r$^\prime$-band over five nights in October 2009. \textit{Bottom of upper panel:} Differential light curve of the comparison stars (comp2/(comp1+comp3)). The change in observed flux for the target is not seen in the light curve of the comparison stars. \textit{Top of lower panel:} FT of SDSS J0005-1002 light curve, where frequencies have been searched up to approximately the Nyquist frequency. The inset figure shows the low frequencies in more detail. The maximum amplitude is measured at a frequency of 0.490103 cycles/d ($P=2.04$ d). The other peaks at low frequencies are aliases due to the window function. The dashed lines indicate the $\sigma$ and $3\sigma$ noise levels. \textit{Bottom of lower panel:} FT of the light curve of the comparison stars. The amplitude for the comparison stars is much smaller than the amplitude in the FT for the target and there are no peaks above the 1$\sigma$ noise level at an amplitude of 13.3\,mma. The amplitude is given in units of milli-modulation amplitude (mma), where 10 mma corresponds to 1\%.}\label{fig:sdss0005_int_lc}
\end{center}
\end{figure}

Figure \ref{fig:sdss0005_int_lc} shows the light curve obtained from the INT data and resulting Fourier transform (FT) for SDSS J0005-1002 and its comparison stars. We measure the maximum amplitude in the FT at a frequency of 0.490103 cycles/d ($P=2.04$ d). We also see alias peaks at low frequencies with comparable amplitude to the main peak (see inset Fig. \ref{fig:sdss0005_int_lc}, \textit{lower panel}) due to the window function. We do not detect any fluctuations in the relative flux of the comparison stars, which is reflected in the small amplitude in the FT.

\begin{figure}
\begin{center}
\includegraphics[width=8cm]{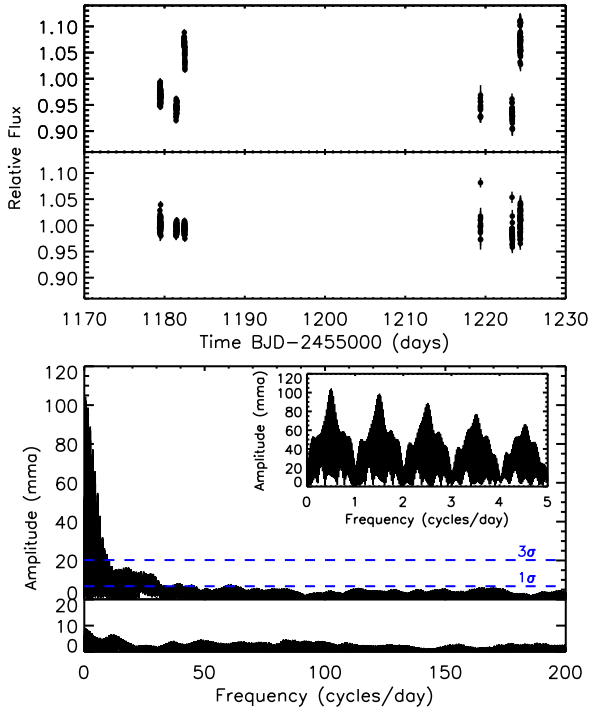}
\caption{\textit{Top of upper panel:} Differential light curve of SDSS J0005-1002 (target/comp3) taken using the STE3 instrument on the SAAO 1.0 m with no filter over four nights in September 2012 and five nights in October 2012. \textit{Bottom of upper panel:} Differential light curve of the comparison stars (comp2/comp3). The scatter on short timescales (i.e. over one night) is comparable between nights. Variations in flux are observed from night-to-night in the light curve of SDSS J0005-1002, while the light curve of the comparison stars is stable. The scatter in the SAAO data is understandably larger than in the INT data, as SDSS J0005-1002 is V=18.3 mag and was observed with a smaller 1.0 m telescope. It also has a smaller field-of-view, thus limiting the number of appropriate comparison stars available for differential photometry. \textit{Top of lower panel:} The corresponding FT of the target light curve, searching up to approximately the Nyquist frequency. The inset figure shows a close-up at low frequencies. The maximum peak is measured at 0.501235 cycles/d ($P=1.99$ d). The other peaks at low frequencies are aliases due to the window function. The dashed lines indicate the $\sigma$ and $3\sigma$ noise levels. \textit{Bottom of lower panel:} FT of the light curve of the comparison stars with no peaks above the 1$\sigma$ noise level at an amplitude of 6.7\,mma.}\label{fig:sdss0005_allsaao_lcft}
\end{center}
\end{figure}

The SAAO light curve and FT for the target and comparison stars are shown in Figure \ref{fig:sdss0005_allsaao_lcft}. As found from the INT light curve, the relative flux of the comparison stars is stable over days. The FT (Fig. \ref{fig:sdss0005_allsaao_lcft}, \textit{top of lower panel}) has a maximum peak at 0.501235 cycles/d ($P=1.99$ d), agreeing approximately with the Fourier analysis of the INT data. 

We also fit the light curves with a sinusoid plus a constant using MPFIT in IDL (\citealt{Markwardt09}) and fold on the best-fitting period. Again, we determine slightly different best-fitting periods for the two data sets, but they agree within error estimates. For the INT data, we find a best-fitting period of $2.104\pm0.030$ days, with a reduced $\chi^2$ of 3.77 ($\chi^2$ of 86.7 over 23 dof). For the SAAO data, we find a best-fitting of $2.110\pm0.001$ days, with a reduced $\chi^2$ of 3.26 ($\chi^2$ of 918.6 over 282 dof). These periods are slightly different from the values determined from the Fourier analysis due to the disparate methods for how the ``best-fitting" periods are calculated. For example, the Fourier analysis method subtracts the mean of the light curve before determining the FT, whereas the sine curve fitting has slightly more flexibility by fitting a sinusoid plus a constant to the data. In addition, the two methods use different criteria to define the ``best'' period, which corresponds to the frequency at the maximum amplitude in the FT, while the ``best'' frequency/period from the sine-fitting is taken at the minimum $\chi^2$ value. 

\begin{figure}
\begin{center}
\includegraphics[width=8cm]{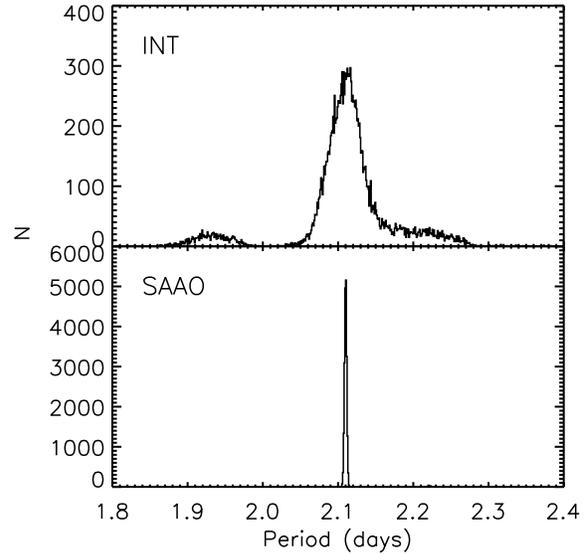}
\caption{Distribution of possible periods for the INT and SAAO data sets after bootstrapping 20,000 times. \textit{Top:} The INT data peaks at a period of 2.110 days with a corresponding 2$\sigma$ uncertainty of 0.045 days, estimated from fitting a Gaussian curve to the peak. \textit{Bottom:} The distribution from bootstrapping the SAAO data peaks at the same period of 2.110 days with a 2$\sigma$ uncertainty of 0.003 days. }\label{fig:P_dist}
\end{center}
\end{figure}

The period uncertainties are independently estimated by bootstrapping the data. Both data sets are fit with a sine wave using MPFIT, then the light curves are resampled by randomly selecting the same number of points and re-fit with a sine wave (``resampling with replacement'', \citealt{Brinkworth05}; \citealt{Diaconis83}). This is repeated 20,000 times. The resultant distribution of possible periods is given in Figure \ref{fig:P_dist}. The distribution of periods for the INT data peaks at 2.110 days with a 2$\sigma$ uncertainty of 0.045 days, while the SAAO data distribution of periods peaks at the same period with a 2$\sigma$ uncertainty of 0.003 days. 

\begin{figure}
\begin{center}
\includegraphics[width=8cm]{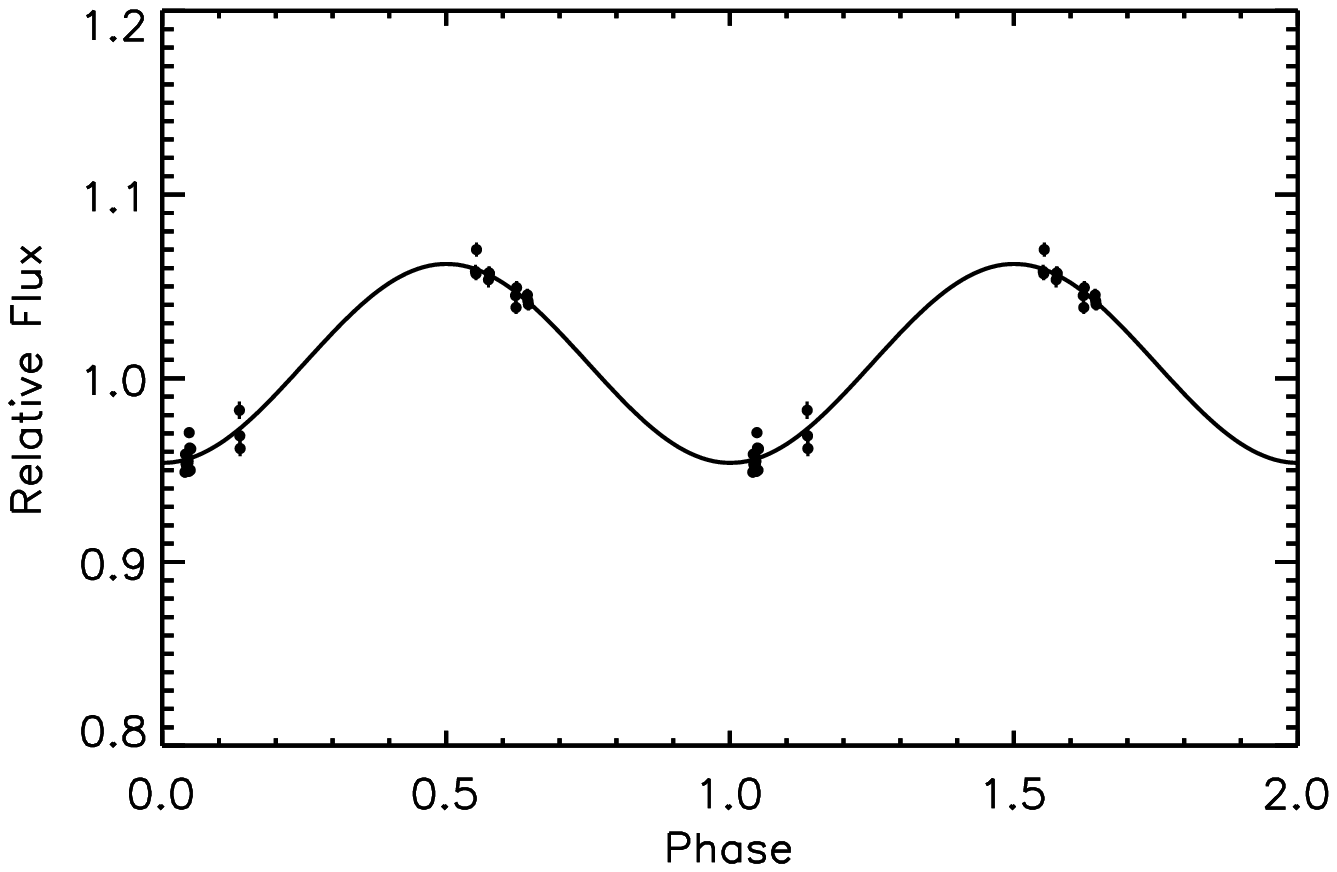}
\includegraphics[width=8cm]{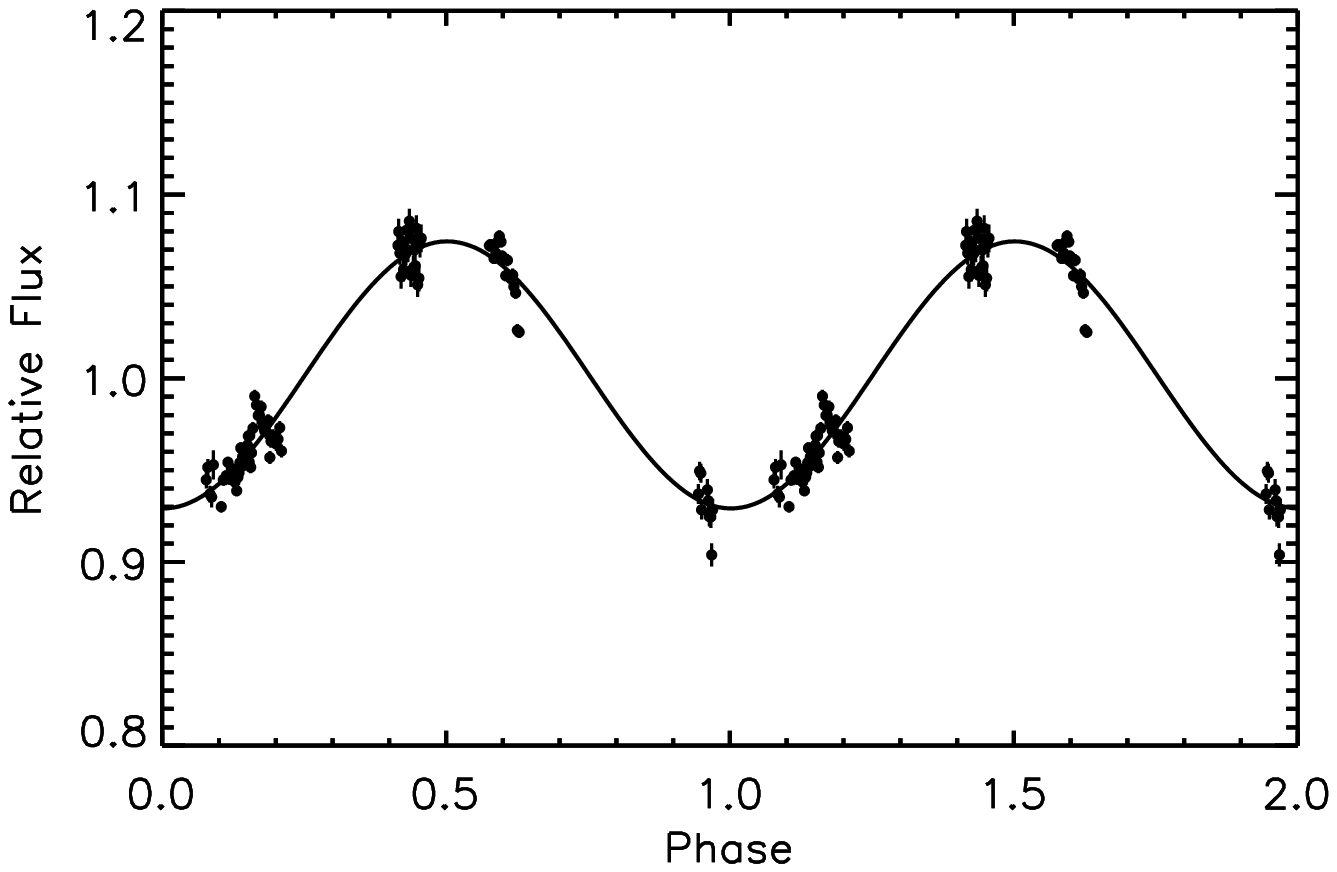}
\caption{\textit{Top:} INT light curve folded on 2.110 days with a starting time of 2455122.1804753 BJD at minimum flux. We find the best-fitting sine curve has a reduced $\chi^2$ of 3.61 and a peak-to-peak amplitude of 10.9\%. \textit{Bottom:} The SAAO light curve folded on the same period and binned by a factor of 2. It has a reduced $\chi^2$ of 3.26 and a peak-to-peak amplitude of 14.7\%. The $\approx$2 day period has made it difficult to observe all phases of the variability with rotation.} \label{fig:wd0003_all_lcfold}
\end{center}
\end{figure}

In Figure \ref{fig:wd0003_all_lcfold}, both sets of light curves are folded on the 2.110 day period. The SAAO light curve is folded on the ephemeris for the time at minimum flux, 
\begin{eqnarray}
BJD = 2456179.1036(48) + 2.110(45)E . \nonumber
\end{eqnarray}
However, this is not used to fold the INT data set, as the period estimate is not accurate enough to link the two data sets, which are separated by nearly three years. Unfortunately, we have not been able to obtain complete coverage over all phases due to the 2 day timescale of variability. As a result, we cannot definitively say whether the photometric variations are sinusoidal or not. 

The amplitude of the INT folded light curve is $10.9\pm0.3$\%, which is not the same as the amplitude of the SAAO folded light curve at $14.7\pm0.2$\%. This is not surprising as the SAAO data was taken without a filter and the INT data was taken in the $r^{\prime}$-band. Photometric variability in magnetic white dwarfs is known to exhibit a wavelength dependence due to spectroscopic variations in the presence of a changing magnetic field configuration (e.g. RE\,J0317-853, \citealt{Vennes03}). 

\begin{figure*}
\begin{center}
\includegraphics[width=18cm]{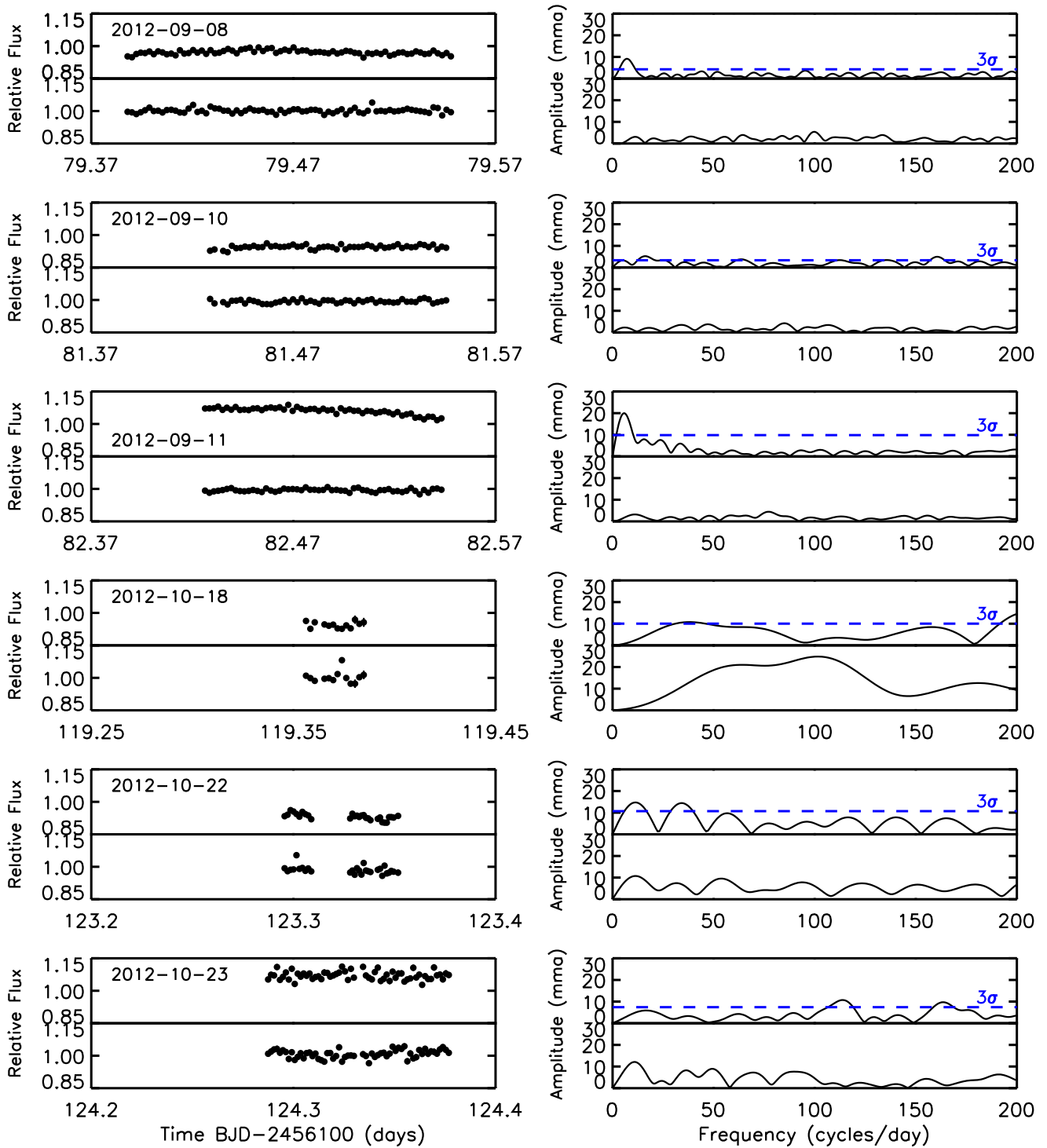}
\caption{Light curves from the individual SAAO nights (\textit{left}) and their corresponding FTs (\textit{right}). Each panel shows the results for the target (\textit{top}) and for the comparison stars (\textit{bottom}). The relative flux for the target is calculated as the target flux divided by the comp3 flux, while the comparison star relative flux is determined as comp2 flux/comp3 flux. We find no evidence for short period fluctuations above a $3\sigma$ detection limit (three times the noise level $\sigma$, dashed line) at an amplitude of $\lesssim$$\pm$0.5\% for the two best light curves taken on 2012-09-08 and 2012-09-10. The light curve of SDSS J0005-1002 taken on 2012-09-11, showing a decrease in flux towards the end of the run, is most likely due to deteriorating sky conditions. Some scatter is seen in the light curves. However, these features on short timescales are also evident for the comparison stars. The peaks at small frequencies in the FTs correspond to the length of the observations.}\label{fig:sdss0005_saao_lcft}
\end{center}
\end{figure*}

The nightly SAAO light curves are not corrected for differential refraction effects due to changes in the airmass during observing because we did not want to unintentionally remove any real long term changes in the flux. The nightly SAAO light curves are shown in Figure \ref{fig:sdss0005_saao_lcft} for the target and comparison star. The flux in the nightly light curves are consistently stable and do not show the secular change in flux with time, indicative of residual atmospheric effects. 

Since other magnetic hot DQ white dwarfs show short term fluctuations on timescales of 210--1050 s (\citealt{Barlow08}; \citealt{Dunlap10}; \citealt{Dufour11}), we analyse the nightly SAAO light curves, each of which are up to 3 hours long (see Table \ref{tab:obs_log}), for short period modulations. The individual light curves of the target and comparison stars, and corresponding FTs are shown in Figure \ref{fig:sdss0005_saao_lcft}. We see some scatter on short timescales in the relative flux, but these features are also evident in our analysis of the comparison stars. The peaks in the FTs at low frequencies correspond approximately to the length of the nightly light curves. A model light curve is generated using the ephemeris for the same times as the SAAO data set, showing the model is consistent with the nightly SAAO light curves in Figure \ref{fig:sdss0005_saao_lcft}. 

To confirm whether any peaks in the individual FTs are real, we determine false alarm probabilities (FAPs) for each of the SAAO data sets using the method in \cite{Alcock00} and \cite*{Kovacs02}. The significance $Sg$ of the highest peak in the periodogram is calculated using, 
\begin{equation}
Sg = \frac{A_{max} - <A>}{\sigma_A},
\end{equation}
where $A_{max}$ is the amplitude $A$ at the highest peak in the FT, $<A>$ is the average amplitude and $\sigma_A$ is the standard deviation of $A$ for the given frequency range. This procedure is carried out for 1000 fake light curves which are generated by randomly shuffling the target light curve and repeating the periodogram analysis. A probability distribution function (PDF) is then calculated from the simulated light curves, where the FAPs are determined for each of the nightly SAAO light curves as 0.059, 0.424, 0.030, 0.434, 0.295 and 0.347 respectively. These are all above a FAP threshold of 0.01 (a 99\% significance detection limit). Furthermore, we detect no significant peaks in the FTs in Figure \ref{fig:sdss0005_saao_lcft} above a 3$\sigma$ detection limit (three times the noise level $\sigma$, dashed line in Fig. \ref{fig:sdss0005_saao_lcft}). Consequently, we conclude that we do not find evidence for photometric variability in SDSS J0005-1002 on a timescale less than 3 hours (also found by K. Williams \& B. Dunlap, priv. comm.) at an amplitude level of $\lesssim$$\pm0.5\%$ ($3\sigma$) for the two best light curves taken on 2012-09-08 and 2012-09-10. At this detection limit, the smallest-amplitude pulsations exhibited by the other variable hot DQs may be undetectable in some of our current SAAO data sets. However, the variable hot DQ pulsations typically have semi-amplitudes of 7 mma (0.7\%) in the optical, with SDSS J1426-5752's main pulsation period having the highest amplitude at 17.5 mma (1.75\%, \citealt{Montgomery08}). These pulsation semi-amplitudes would have just been detectable in the two best SAAO data sets.

%%%%%%%%%%%%%%%%%%%%%%%%%%% DISCUSSION %%%%%%%%%%%%%%%%%%%%%%%%%%%%%%%%
\section{Discussion}

\subsection{The variability is due to rotation, not pulsations.}
We have discovered the first long period photometric variations for a hot DQ white dwarf, SDSS J0005-1002, with a period of $2.110\pm0.045$ days and peak-to-peak amplitude of 11\%. We also find no evidence for fluctuations on timescales of less than a few hours at an amplitude level of $\lesssim$$\pm$0.5\%. In contrast, the other variable hot DQs show short term fluctuations up to $\sim$1000\,s. The vast majority of pulsating white dwarfs have modes shorter than 2000\,s (e.g. \citealt*{Winget08}). The longest pulsation mode ever measured is 4444\,s (\citealt{Hermes12}), but this is for a rare extremely low-mass white dwarf ($M\sim0.17M_{\odot}$). We believe the variability seen in SDSS J0005-1002 is therefore due to rotation, and not pulsations, as no white dwarf has ever been observed to pulsate with modes of days. 

The spin period for SDSS J0005-1002 is consistent with rotation period measurements for pulsating white dwarfs, which are typically around a day and determined from the splitting of their pulsation modes (e.g. \citealt{Kawaler04}; \citealt{Winget08}). Approximately 40\% of magnetic white dwarfs show photometric variations with rotation, and the majority have spin periods of hours to a few days (Brinkworth et al. 2013, submitted; Lawrie et al. 2013, in prep). 

The photometric variations may be due to Zeeman split\,/\,broadened lines changing in strength and position with rotation from a varying magnetic field strength and configuration across the whole stellar surface. In a carbon-dominated hot DQ white dwarf, such as SDSS J0005-1002, the broad, Zeeman split C\,{\sc ii} atomic feature is likely changing in strength across the surface, hence causing the significant flux variations. Alternatively, star spots may be present on the surface of SDSS J0005-1002 if the atmosphere is, at least, partially convective. In the case of the cool, isolated DA white dwarf, WD 1953-011, a field-enhancement spot covering $\sim$10\% of the surface is thought to cause sinusoidal variations of $\approx$2\% with spin on a period of 1.44 days (\citealt{Maxted00}; \citealt{Brinkworth05}). High-resolution, time-resolved spectroscopic observations of SDSS J0005-1002 over the rotation period are required to investigate how the spectral features change with the rotational phase and to determine the possible cause behind the photometric variations. 

\subsection{Are all hot DQ WDs magnetic?}
Up to 70\% of hot DQ white dwarfs are magnetic (\citealt{Dufour10}; \citealt{Dufour13}). This incidence of magnetism is much higher than for the general white dwarf population, perhaps suggesting that all hot DQs are magnetic. \cite{Kepler13} recently found 521 DA white dwarfs with detectable Zeeman split lines, with magnetic fields ranging from $\sim$1 to 733\,MG, in the SDSS DR7 white dwarf catalogue (\citealt{Kleinman13}), equating to $\sim$4\% of all DA white dwarfs observed. The fraction of white dwarfs with weak magnetic fields is thought to be higher. \cite{Jordan07} estimated $11-15$\% have kilo-Gauss field strengths, although a re-analysis of the white dwarfs with kilo-Gauss field strengths by \cite{Landstreet12} determined that 10\% of white dwarfs have magnetic fields. However, \cite{Kawka12} found $5\pm2$\% have field strengths $\lesssim$100\,kG. By contrast, \cite{Kawka07} estimated the incidence of magnetism in the Solar neighbourhood (within 13\,pc of the Sun) as $21\pm8$\%, and \cite{Lieb03} and \cite{Giammichele12} found $\approx$10\% of white dwarfs within 20\,pc of the Sun have magnetic fields.  

Interestingly, no magnetic field has ever been detected in a pulsating hydrogen-dominated atmosphere DA white dwarf, despite several hundred DA magnetic white dwarfs now known, whereas magnetic fields are measured in some pulsating carbon-dominated hot DQ white dwarfs. 

Although \cite{Montgomery08} predicted that the prototype hot DQ variable SDSS J1426-5752 should indeed pulsate, they found that the observed pulse shape was different to that seen for normal DA white dwarf pulsators, having a flat maximum and sharp minimum. Large amplitude pulsating DA white dwarfs are typically characterised by the opposite behaviour, a flat minimum and sharp maximum. This unusual pulse shape is also seen for SDSS J2200-0741 (\citealt{Barlow08}; \citealt{Dufour09}) and SDSS J1337-0026 (\citealt{Dunlap10}, although \citealt{Dufour11} did not find this in their FUV light curves). Since these stars also have magnetic fields\footnote{\cite{Dufour13} recently detected a magnetic field for SDSS J1337-0026.}, \cite{Green09} and \cite{Dufour09} suggested that the different pulse shape could be due to the presence of the magnetic field. Furthermore, another hot DQ SDSS J2348-0942 has no known magnetic field and a sinusoidal pulse shape, suggesting the presence of magnetism may influence the observed pulse shape in magnetic DQs. 

\cite{Williams13} recently announced the discovery of a photometrically variable ``warm'' DQ magnetic white dwarf, SDSS J1036+6522, thought to be a transition object between the hot, carbon-dominated DQs and the cool, helium-dominated DQs. SDSS J1036+6522 has a mean magnetic field strength of 3 MG and is cooler than the hot DQs with an estimated effective temperature of $T_{\rm eff}$\,$\approx$\,15,500\,K. The star displays photometric monoperiodic modulations on a period of 1115\,s at a small amplitude of 0.44\% with a sinusoidal pulse shape. Since this period is similar to ZZ Ceti and hot DQ pulsation modes, \cite{Williams13} conclude that SDSS J1036+6522 is also most likely pulsating. Even though SDSS J1036+6522 has many similar characteristics to the hot DQs, it contrasts with previous findings of hot DQs which either appear to be magnetic and have unusual, asymmetric pulse shapes or be non-magnetic and have sinusoidal pulses. Thereby, this illustrates that if the mechanism causing the photometric variability in SDSS J1036+6522 is the same as the variable hot DQs, then the pulse shape is not an indicator for the presence or absence of a magnetic field. 

\subsection{Have pulsations in some hot DQs been mistaken for rotation of a magnetic WD?}
The pulsations interpretation for the short term variable hot DQs is probably real in most cases. However, not all of them exhibit multi-periodic modes in their FTs, a characteristic indicative of pulsators, and therefore the single mode pulsators may actually be photometrically variable due to rotation. Rapid rotation periods as short as tens of minutes have been measured for RE J0317-853 at 725\,s (\citealt{Barstow95}; \citealt{Ferrario97}) and SDSS J2257+0755 at 1354\,s (Lawrie et al. 2013, in prep), comparable to the length of pulsation modes. Both of these DA white dwarfs have effective temperatures $T_{\rm eff} > 30,000$~K well beyond the hydrogen instability strip, and therefore these are not pulsation modes. Similarly, the ``warm" DQ, SDSS J1036+6522, may in fact be rotating with a period of 1115\,s (\citealt{Williams13}) and not necessarily a pulsator. We suggest that all hot DQ white dwarfs should be observed for both long and short period photometric variability. This can also be used as a method for indicating whether some of the other hot DQs may be magnetic. For example, a hot, variable white dwarf was detected in the \textit{Kepler} field by \cite*{Holberg11} with photometric modulations of $\approx$5\% peak-to-peak on a period of 6.1375\,h. Subsequent high signal-to-noise spectra confirmed that the star was in fact magnetic. 

\subsection{On the origin of the magnetic field in hot DQ WDs.}
Magnetism in white dwarfs, in general, is thought to originate from either a magnetic main-sequence progenitor star or generate in a binary merger, and therefore, magnetism is expected to be found at all points along the white dwarf cooling tracks. Hot DQs may descend from hydrogen and helium deficient pre-white dwarf stars (similar to the helium-atmosphere white dwarfs), and as they cool, any remaining helium in the atmosphere diffuses to the outer layer, where it appears as a helium-dominated DO/DB white dwarf (\citealt{Dufour08b}; \citealt{ALT09}). When the convection zone develops (at $T_{\rm eff}\sim24,000$~K, \citealt{Dufour08b}), the thin helium layer gets diluted and the star transforms into a carbon-dominated hot DQ white dwarf. If the hot DQs are linked with the previously known cooler DQs in an evolutionary sequence (as suggested by \citealt{Dufour08b}; \citealt{Dufour13}), one would expect to find the same high incidence of magnetism in the cooler helium-dominated atmosphere DQ white dwarfs as observed for the hot DQs, which does not appear to be the case (\citealt{Dufour13}). Perhaps the magnetic field detected in hot DQs is generated in the developing carbon-oxygen convection zone, as the star converts from a DB to a hot DQ, rather than being a fossil field from the progenitor main-sequence star or created during the common envelope phase. Subsequently, as the star cools further and the convection reduces, the magnetic field dies, explaining the absence of magnetism in the cooler DQs. 
%There is some evidence that the sequence of DQ WDs may be more massive than most white dwarfs (e.g. \citealt{Dufour13}), possibly suggesting they evolve from more massive progenitor stars. 

\section{Conclusions}
In summary, hot DQ white dwarfs remain enigmatic objects. Most, if not all, are magnetic, in contrast to the white dwarf population in general, and many appear to pulsate. We have shown here that one also photometrically varies as it rotates on a period of $2.110\pm0.045$ days. We suggest that all hot DQs should be observed for long period modulations, indicative of rotation, and that some hot DQ ``pulsators", especially those with a single oscillation mode, should be observed to test the possibility that some short term variables may be rotators after all. The variation may be due to star spots in a convective atmosphere, or changes in the Zeeman splitting and line strengths due to a varying field strength and configuration across the surface of the star. Therefore, these stars, and SDSS J0005-1002 in particular, should be targeted for high-resolution, time-resolved spectroscopic observations to investigate how the spectral features change with the rotational phase, and although more difficult to obtain, time-resolved spectropolarimetry over the rotation period would provide a unique insight into a possibly changing magnetic field and the cause of the fluctuating brightness. Magnetism may play a key role, or provide clues, to the origin and evolution of hot DQ white dwarfs.

%%%%%%%%%%%%%%%%%%%%%%%%%%%%%%%%%%%%%%%%%%%%%%%%%%%%%%%%%%%%%%%%%%%%%%%%%%%%%%
%% Acknowledgements                                                         %%
%%%%%%%%%%%%%%%%%%%%%%%%%%%%%%%%%%%%%%%%%%%%%%%%%%%%%%%%%%%%%%%%%%%%%%%%%%%%%%
\section*{Acknowledgements}
Based on observations made with the INT operated on the island of La Palma by the Isaac Newton Group in the Spanish Observatorio del Roque de los Muchachos of the Instituto de Astrofisica de Canarias. This paper uses observations made at the South African Astronomical Observatory (SAAO). This research has made use of the SIMBAD database, operated at CDS, Strasbourg, France.

%%%%%%%%%%%%%%%%%%%%%%%%%%%%%%%%%%%%%%%%%%%%%%%%%%%%%%%%%%%%%%%%%%%%%%%%%%%%%%
%% Bibliography                                                             %%
%%%%%%%%%%%%%%%%%%%%%%%%%%%%%%%%%%%%%%%%%%%%%%%%%%%%%%%%%%%%%%%%%%%%%%%%%%%%%%
\bibliographystyle{mn2e} 
\bibliography{hotDQ_sdss0005_revised}

\end{document}